\documentstyle[emulateapj,onecolfloat]{article}
\input epsf

\newcommand{\simlt}{\lesssim}
\newcommand{\simgt}{\gtrsim}
\newcommand{\zem}{z_{\rm s}}
\newcommand{\aem}{a_{\rm s}}
\newcommand{\clk}{C_\ell}
\newcommand{\smooth}{\sigma}

\begin{document}

\twocolumn[
\title{A New Algorithm for Computing Statistics of Weak Lensing by
Large-Scale Structure}
\author{Martin White and Wayne Hu}
\affil{
Harvard-Smithsonian Center for Astrophysics, Cambridge, MA 02138\\
       Institute for Advanced Study, Princeton, NJ 08540}
\authoremail{mwhite@cfa.harvard.edu}
\authoremail{whu@sns.ias.edu}

\begin{abstract}
\noindent
\rightskip=0pt
We describe an efficient algorithm for calculating the statistics
of weak lensing by large-scale structure based on a tiled set of
independent particle-mesh N-body simulations which telescope in resolution
along the line-of-sight.  This efficiency allows us to predict not 
only the mean properties of lensing observables such as the
power spectrum, skewness and kurtosis of the convergence, but also
their sampling errors for finite fields of view, which are themselves 
crucial for assessing the
cosmological significance of observations.
We find that the nongaussianity of the distribution substantially
increases the sampling errors for the skewness and kurtosis
in the several to tens of arcminutes 
regime, whereas those for the power spectrum are only fractionally
increased even out to wavenumbers where shot noise from the intrinsic
ellipticities of the galaxies will likely dominate the errors. 
\end{abstract}
]
\section{Introduction} \label{sec:intro}

Weak lensing of background galaxies by foreground large-scale structure
offers an opportunity to directly probe the mass distribution on large
scales over a wide range of redshifts.
In this paper we describe an N-body based algorithm optimized for weak
lensing calculations which can be run on workstation-class computers.
The method is fast and efficient, allowing the exploration of parameter
space and production of many realizations of a given model to assess the
statistical significance of any result.

Weak lensing of distant galaxies by large scale structure shears and
magnifies their images.  As first pointed out by
Blandford et al.~(\cite{Blaetal91}) and Miralda-Escude~(\cite{Mir91}),
these effects are of order a few percent in adiabatic cold dark matter
models making their observation challenging but feasible.   
Early predictions for the power spectrum of the shear and convergence were
made by Kaiser~(\cite{Kai92}) on the basis of linear perturbation theory.
Likewise the skewness of the convergence in perturbation theory was computed
by Bernardeau, van Waerbeke \& Mellier (\cite{BerWaeMel97}).
Jain \& Seljak~(\cite{JaiSel97}) estimated the effect of
non-linearities in the density through analytic fitting formulae
(Peacock \& Dodds~\cite{PeaDod96}) and showed they substantially increase
the power in the convergence below the degree scale. 

On subdegree scales, a full description of weak lensing therefore requires
numerical simulations, the most natural being N-body simulations.
N-body codes are ideally suited for weak lensing calculations since on the
relevant scales only gravity is involved, bypassing the need for a
treatment of hydrodynamic and radiative transfer effects.  
The evolution of density perturbations into the non-linear regime by N-body
techniques is now a well-developed field.
The particle-mesh (PM) N-body technique provides an efficient means of
simulating the evolution of structure.  
Its speed makes it ideal for the rapid exploration of cosmological models
and the calculation of statistical properties of the lensing observables,
e.g.\ the sampling variance on estimators of the power spectrum, skewness
and kurtosis of the convergence.
While Lagrangian perturbation theory is arguably even more efficient
(Waerbeke, Bernardeau \& Mellier~\cite{WaeBerMel98}),
without the proper non-linear dynamics one cannot guarantee that the statistics
are faithfully reproduced.
 
The main drawback of the PM technique is the lack of {\it angular\/} dynamic
range, due partially to the broad kernel that describes the efficiency with
which structures along the line-of-sight lens the sources
(Jain, Seljak \& White \cite{JaiSelWhi98}). 
We show here that this problem may be in large part overcome by tiling the
line-of-sight with simulations of increasing resolution. 

The lensing signal is calculated by ray-tracing through the simulations
(Blandford et al.~\cite{Blaetal91},
Wambsganss, Cen \& Ostriker~\cite{WamCenOst98},
Jain et al.~\cite{JaiSelWhi98};
Couchman, Barber \& Thomas~\cite{CouBarTho98};
Fluke, Webster \& Mortlock~\cite{FluWebMor98};
Hamana, Martel \& Futamase~\cite{HamMarFut99}).
In the weak lensing regime, a key simplification is that one can use
unperturbed photon paths to perform the relevant line-of-sight integrals,
eliminating the need for explicit ray tracing  
(Blandford et al.~\cite{Blaetal91}; Hui, private communication;
see e.g.~Bartelmann \& Schneider~\cite{BarSch} for a discussion).
This allows one to incorporate the lensing right into the time evolution of
the code, eliminating the need to output the density field along the way and
allowing very dense sampling of the integrals.
While the evaluation of the convergence along an unperturbed path is
self-consistent within the framework of weak lensing, the results 
must be checked against a full ray tracing method.
The simulations reported in Jain et al.~(\cite{JaiSelWhi98}) suggest
that the approximation is good to $10^{-3}$ in power for models similar to
the one reported in this paper.

In this paper we shall concentrate on a specific cosmological model.
It is a cosmological constant cold dark matter model ($\Lambda$CDM) with
$\Omega_m=0.3=1-\Omega_\Lambda$, a scale-invariant spectrum of adiabatic
perturbations ($n=1$) with a matter power spectrum described by the fitting
function of Bardeen et al.~(\cite{BBKS}) with $\Gamma_{\rm BBKS}=0.2$.
The model is normalized to the {\sl COBE\/} 4-year data using the method
of Bunn \& White~(\cite{BunWhi}).  This corresponds to $\sigma_8=1.2$,
slightly above that inferred from the abundance of rich clusters
(Eke et al.~\cite{EkeColFreHen}, Viana \& Liddle~\cite{ViaLid}).

The outline of the paper is as follows.  In \S\ref{sec:code} we describe
our implementation of a PM code and lensing evaluation.  In \S\ref{sec:tiling}
we introduce the tiling technique.  We present results for the power spectrum
of the convergence and sampling errors in its estimation in \S\ref{sec:power}
and analogous results for the skewness and kurtosis of the convergence
\S\ref{sec:skewness}.  A comparison of our tiling method and those based
on single simulations is presented in \S\ref{sec:singlebox}.
We conclude in \S\ref{sec:discussion}.

\section{The PM-Lensing Code} \label{sec:code}

To evolve the dark matter distribution in the non-linear regime, we use a
particle-mesh (PM) code described in detail in
Meiksin et al.~(\cite{MWP}) and White~(\cite{W99}).
The simulations reported here use either $128^3$ or $256^3$ particles and
a $256^3$ or $512^3$ force ``mesh''.
The initial conditions are generated by displacing particles from a regular
grid using the Zel'dovich approximation.
The simulations are started at $1+z=35$ and evolved to the present ($z=0$)
using adaptive steps in the log of the scale factor $a=(1+z)^{-1}$.
The force on each particle is calculated from the density using Fourier
Transform (FT) techniques with a kernel $-\vec{k}/k^2$. The gridded fields
are computed from the particle data using CIC charge assignment
(Hockney \& Eastwood~\cite{HocEas}).
The time step is dynamically chosen as a small fraction of the inverse square
root of the maximum acceleration, with an upper limit of $\Delta a/a=3$ per
cent per step.  The code typically takes 200--300 time steps between $1+z=35$
and $z=0$.

The new ingredient, beyond simple N-body evolution of the density field, is
the calculation of the convergence along a bundle of rays.
In the weak lensing approximation, calculation of this scalar quantity in
any field is sufficient to enable calculation of all of the other quantities
(e.g.~the shear $\vec{\gamma}$).
We assume here for simplicity that the sources all lie at one redshift
$\zem=1$.  The code as written allows multiple source redshifts,
but we restrict ourselves to the single source plane in this paper.

Before the N-body evolution begins, we generate geo\-desics, in code coordinates,
for $N_{\rm los}=128^2$ or $256^2$ lines-of-sight by integrating
\begin{equation}
  {dD_{||}\over da} = {1 \over a^2 H(a)} \,,
\end{equation}
where $D_{||}$ is the comoving distance parallel to the line-of-sight.
The lines-of-sight originate in a square lattice at $\zem$ and
converge upon an observer situated at the center of one face of the box at
$z=0$.  We further demand that the field of view never subtend more than
a box length to avoid introducing artifacts due to periodic boundary
conditions.
We make the small angle approximation and assume that $D_{||}$ lies parallel
to the $z$-axis for all rays, thus the coordinates perpendicular to the
line-of-sight scale linearly\footnote{This is appropriate for the flat
universes we deal with in this paper.  In a curved universe, the angular
diameter distance must be used to calculate the ``opening distance'' of any
ray from the center of the box as a function of redshift.} with $D_{||}$.


The N-body code is then run, and once the evolution reaches a redshift of
$\zem$, we integrate the lensing equation
\begin{equation}
  \kappa(\vec{x}_\perp) = D_{\rm s} \int 
    d D_{||}\ t(1-t) \nabla_\perp^2 \Phi(\vec{x})
\label{eqn:losint}
\end{equation}
in addition to the gravitational force.
Here $t\equiv D(a)/D_{\rm s}\in [0,1]$ is the dimensionless distance to the
source.  For multiple sources one can replace the kernel for source $i$ with
$t(1-t)$ with $t(t_{{\rm s} i}-t)/t_{{\rm s}i}$ where
$t_{\rm s} = D_{{\rm s}i}/D_{\rm s}$ is the distance to the $i$th source in
units of the distance to the furthest source $D_{\rm s}$.


The source $\nabla_\perp^2 \Phi(\vec{x})$ is calculated in the box using FT
methods under the small-angle approximation.
The particles are assigned to the nearest point on a grid
(NGP; Hockney \& Eastwood~\cite{HocEas}) to obtain the density distribution.
The FT of this distribution, $\delta_k$, is then multiplied by
$-{3\over 2}\Omega_m H_0^2 a^{-1} k_\perp^2/k^2$ 
and the transform inverted.
Within each time step, we assume that the potential is slowly varying
$\Phi(a+\delta a)\approx \Phi(a)$.
Since time steps are separated by $\Delta a/a\sim 0.01$, much less than the
expansion time on which $\Phi$ varies, this is a very good approximation.
The integral is evaluated by taking $N$ points along each line-of-sight and
time step assuming the potential is frozen.
By increasing $N$, we find that $N\sim 10^2$ dynamically chosen points
suffices for convergence. 
This sub-step integration range runs from the $a$ of the last time step in the
code to the present $a$. The integral in Eq.~(\ref{eqn:losint}) is therefore
densely sampled and the $\kappa$ correctly evaluated at the $a$ corresponding
to the box redshift.

\begin{table}
\begin{center}
\caption{\label{tab:tile}}
{\sc Tiling Solution\\}
\begin{tabular}{cc|cc}
\tablevspace{3pt}
\hline
$a_{\rm out}$ & $L_{\rm box}$ &
$a_{\rm out}$ & $L_{\rm box}$ \\
\hline
0.537& 245& 0.822&  75\\
0.577& 245& 0.841&  75\\
0.610& 195& 0.861&  75\\
0.646& 195& 0.881&  75\\
0.675& 155& 0.902&  75\\
0.707& 155& 0.924&  75\\
0.732& 120& 0.946&  75\\
0.759& 120& 0.970&  75\\
0.780&  95& 0.994&  75\\
0.803&  95& 1.000&  75\\
\hline
\end{tabular}
\end{center}
\footnotesize
NOTES.---%
The tiling solution for our $\Lambda$CDM model assuming a single source
redshift $\zem=1$, i.e.~$\aem=0.5$, that uses 6 box sizes and 20 tiles.
The column $a_{\rm out}$ gives the scale-factor at which each tile is output.
(A tile contains that part of the integration of $\kappa$ lying between the
last output and $a_{\rm out}$.) The size of the simulation box used for that
tile (in $h^{-1}\,$Mpc) is also given.
\end{table}

We first test the Limber approximation (see Appendix) which says that only
modes perpendicular to the line-of-sight contribute to the integral in
Eq.~(\ref{eqn:losint}).
In this approximation, the 2 dimensional Laplacian can be replaced with
a 3 dimensional Laplacian which in turn can be expressed in terms of
the density perturbation through the Poisson equation:
\begin{equation}
  \kappa(\vec{x}_\perp) = {3 \over 2}\Omega_m H_0^2 D_{\rm s} \int 
  d D_{||}\ t(1-t) [\delta(\vec{x})/a].
\label{eqn:losintdelt}
\end{equation}
Using integration by parts the error induced by this replacement should
be ${\cal O}(\Phi)\sim 10^{-5}$ 
(Jain, et al.~\cite{JaiSelWhi98}).

We have run a $256^3$ PM simulation of our $\Lambda$CDM model in a
$125h^{-1}$Mpc box using Eqs.~(\ref{eqn:losint}) and (\ref{eqn:losintdelt})
to compute $\kappa$ in a grid of $256^2$ lines-of-sight.
The two track each other very well.
The power spectra computed from the two fields are almost identical, as are
the histograms of $\kappa$.
In a line-of-sight by line-of-sight comparison 
Eqs.~(\ref{eqn:losint}) and (\ref{eqn:losintdelt}) return values for $\kappa$
that deviate by at most $0.03$ and on average (rms) $0.003$.
For comparison, the rms fluctuation on the grid scale in these planes is
nearly an order of magnitude larger than this: $\sigma_\kappa\simeq 0.02$.
Some of this scatter is no doubt induced by our small-angle approximation in
computing $\nabla_\perp^2$, while some comes from the finite size of the box.
Since the integration has traced across the box 19 times and at each edge we
can pick up a term ${\cal O}(\Phi)$, this level of variance agrees roughly
with our expectations.
In the absence of finite box size effects, we expect Eqs.~(\ref{eqn:losint})
and (\ref{eqn:losintdelt}) would match even more closely 
and there is reason to believe that our evaluation of
Eq.~(\ref{eqn:losintdelt}) is the more accurate 
(see also Stebbins~\cite{Ste}).

Since Eq.~(\ref{eqn:losintdelt}) is less computationally expensive, 
we will
adopt it from this point on.
The fact that this approximation works well shows that the integral
Eq.~(\ref{eqn:losintdelt}) is sensitive only to those modes in the box which
are perpendicular to the line-of-sight.  This is an important point to remember
when considering questions of sample- or run-to-run variance.

Finally the entire bundle of rays is rotated at a
random angle to the box faces and placed at a random offset from the box center.
This ensures that the rays do not trace parallel to the edges of the
simulation box and the grid used to define the density.
Since the simulation uses periodic boundary conditions, we actually
compute the density in a periodic universe.  Thus each time a ray leaves the
box it is remapped into it using periodicity.

\begin{table}
\caption{\label{tab:numsim}}
\begin{center}
{\sc Number of Simulations\\}
\begin{tabular}{ccc}
\tablevspace{3pt}
\hline
$L_{\rm box}$ & $N_{256}$ & $N_{512}$ \\
\hline
245 & 76 & 10 \\
195 & 20 & 10 \\
155 & 20 & 10 \\
120 & 20 & 10 \\
 95 & 21 & 10 \\
 75 & 36 & 15 \\
\hline
\end{tabular}
\end{center}
\footnotesize
NOTES.---%
The sizes of the simulation boxes used (in $h^{-1}\,$Mpc) and the
number of independent boxes of resolution of that size, with 
both $256^3$ and $512^3$ mesh resolutions.
\end{table}

\begin{figure}
\begin{center}
\leavevmode
\epsfxsize=3.5in \epsfbox{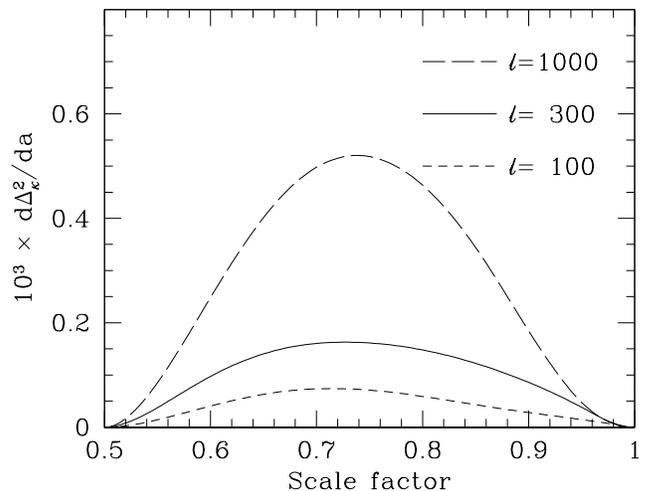}
\end{center}
\caption{\footnotesize%
The contribution to $\Delta_\kappa^2$ as a function of scale-factor for
$\ell=100$, 300, 1000 from Eq.~(\protect\ref{eqn:semianalytic}).}
\label{fig:dclda}
\end{figure}

\section{Tiling} \label{sec:tiling}

A photon from $z\sim 1$ traverses many Gpc on its way to us 
whereas the large-scale structure responsible for lensing 
spans the Mpc range
and below.  Simulating the full range of scales implied 
is currently a practical impossibility.
A solution commonly employed in the literature is to recycle the output of
a single smaller simulation, i.e. sum the contributions of 
the density, projected to the midplane, of the given simulation 
at a series of discrete redshifts.  We propose here a 
``tiling'' alternative
that addresses three potential problems with the traditional technique:
the lack of statistical independence of the fluctuations, the loss of
{\it angular} resolution in the projection, and the discreteness
of the projection.

\begin{figure*}
\begin{center}
\leavevmode
\epsfxsize=3.6in \epsfbox{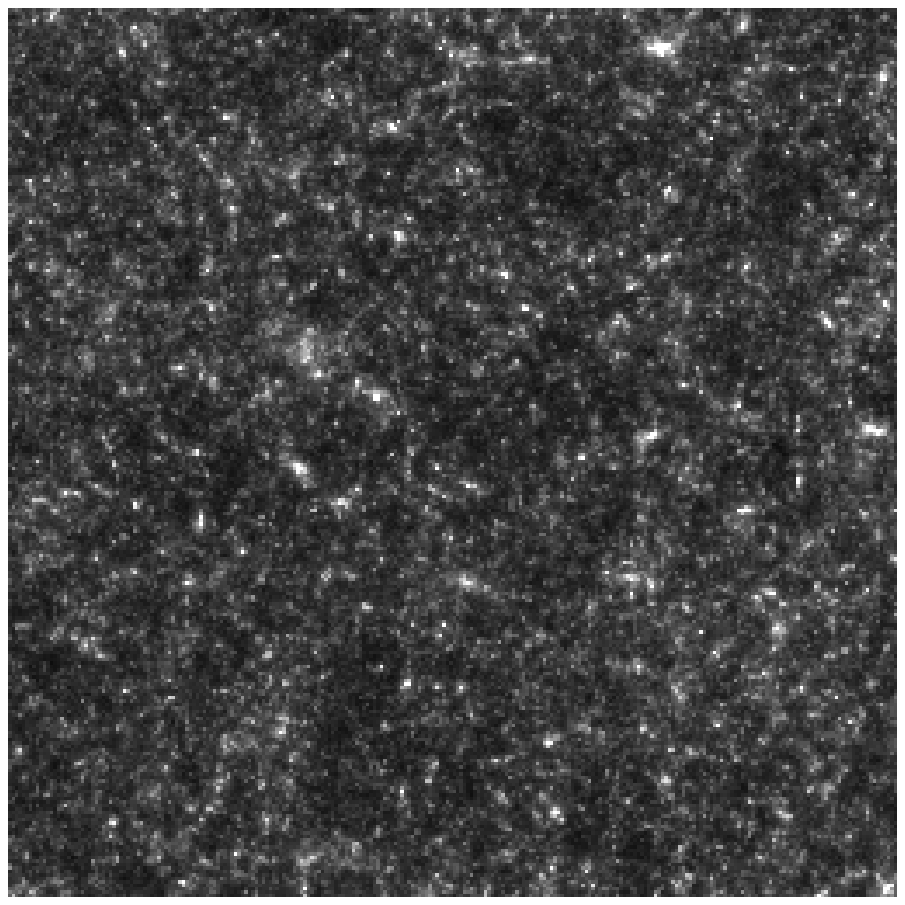}
\epsfxsize=3.6in \epsfbox{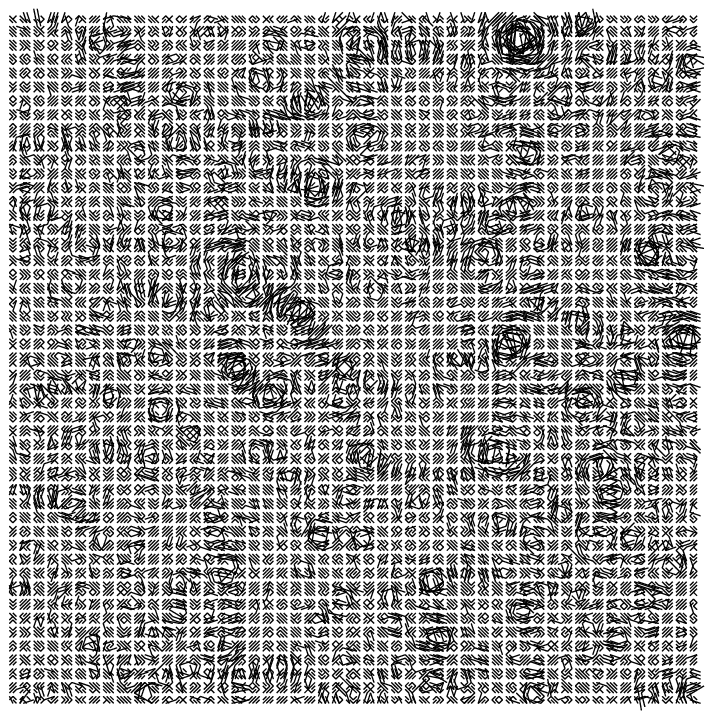}
\end{center}
\caption{\footnotesize%
(left) An image, $6^\circ$ on a side, of the convergence $\kappa$
{}from a single realization of our tiling solution.
The grey scale is linear from $\kappa=-0.05$ to $0.15$.
(right) The shear field, $\gamma_i$, derived from the left panel.
The lines have been exaggerated, the amplitude of the shear is at
the percent level as in (a).}
\label{fig:kapfield}
\end{figure*}

We maintain the the statistical independence of the fluctuations 
by employing multiple independent simulations to tile the line-of-sight.
We are free then to adjust the sizes of the simulation boxes and in
particular can make them  
smaller and smaller as the 
rays converge on the observer (see Table~\ref{tab:numsim}).
Recall that as the lines-of-sight converge,
they probe ever smaller physical separations
for a given angular separation.  
Since the lensing kernel
is so broad, even structure quite close to the observer contributes to the
signal.  In fact, 
the non-linear amplification of structure at low-$z$ implies that
on large angular scales the lensing kernel is skewed toward the 
observer (see Fig.~\ref{fig:dclda}).

Specifically for each simulation box, 
the code outputs the contribution to the $\kappa$ planes
at specified redshifts (see Table~\ref{tab:tile}), typically spaced in $a$
so that the photons traverse the box once between each output.
Note that this is {\it not\/} the same as simply computing the projected
density at the midplane.  The full integral, with the evolution of the
potential and the geometry of the rays etc, is being computed within each
tile.
After each output the offset and random orientation of the rays are chosen
anew and the integration is started afresh for the next segment.

If we simulate only a single box, 
the integral of Eq.~(\ref{eqn:losintdelt}) is simply given by the
sum of the planes from that box. 
However with multiple simulation boxes run with the same tiling scheme,
we can then construct our final $\kappa$ plane as the sum of planes from 
different simulations.  In practice, 
we shrink the box size so that it fits
in the field-of-view at the endplane until it reaches the
non-linear scale.  The non-linear scale must be within the box 
at the relevant epoch to ensure that the PM code evolves
the density correctly.
Nonetheless we shall demonstrate that this box resizing technique is very
effective by comparing results from a series of low resolution ($256^3$)
simulations to our higher resolution ($512^3$) simulations done in a box of
a single size (\S\ref{sec:singlebox}). 

The result at the end of the simulation(s) is a grid $\kappa$ along
lines-of-sight spaced equally in angle.
We then calculate the shear from this grid by using
\begin{eqnarray}
\widetilde{\gamma}_1 &=& {\ell_1^2-\ell_2^2\over \ell_1^2+\ell_2^2}\ \widetilde{\kappa} \,,\\
\widetilde{\gamma}_2 &=& { 2\ell_1 \ell_2  \over \ell_1^2+\ell_2^2}\ \widetilde{\kappa} \,,
\end{eqnarray}
where $\widetilde{\kappa}$ is the 2D FT
of the convergence field,\footnote{Because the field is not
periodic, it is important to zero-pad the FT array before computing
$\widetilde{\kappa}$.} 
and 
${\vec \ell}=(\ell_1,\ell_2)$ is
the Fourier variable conjugate to the position on the sky.

We show in Fig.~\ref{fig:kapfield} the convergence $\kappa$ and the derived
shear field $\gamma_i$, from one of the $512^3$ simulations using the tiling
scheme described in Table~\ref{tab:tile}.  
The field is $6^\circ$ on a side and contains $256^2$ lines-of-sight.
{}From our multiple simulations, we are able to generate many independent
fields of this size and resolution.  In the following sections we discuss
the statistics of these fields based on $512$ random combinations 
of the tiles listed in Table~\ref{tab:numsim} for both the
low and high resolutions sets.

\begin{table*}
\begin{center}
\caption{\label{tab:nbody}}
{\sc Simulation Properties\\}
\begin{tabular}{ccccccc}
\tablevspace{3pt}
\hline
$a_{\rm mid}$ & Weight & $L_{\rm box}$ & $\theta_{\rm box}$ &
$L_{\rm mesh}$ & $\theta_{\rm mesh}$ & $m_{\rm part}$ \\
& & ($h^{-1}$Mpc) & (arcmin) & ($h^{-1}$kpc) & (arcmin) & ($10^9M_\odot$) \\
\hline
0.518& 0.05& 245&   385&  479&  0.75& 73\\
0.557& 0.13& 245&   433&  479&  0.85& 73\\
0.593& 0.19& 195&   389&  381&  0.76& 37\\
0.628& 0.22& 195&   438&  381&  0.86& 37\\
0.660& 0.24& 155&   393&  303&  0.77& 19\\
0.691& 0.25& 155&   444&  303&  0.87& 19\\
0.719& 0.25& 120&   388&  234&  0.76& 8.6\\
0.745& 0.25& 120&   438&  234&  0.85& 8.6\\
0.769& 0.23&  95&   391&  186&  0.76& 4.3\\
0.792& 0.22&  95&   441&  186&  0.86& 4.3\\
0.812& 0.20&  75&   394&  146&  0.77& 2.1\\
0.831& 0.19&  75&   444&  146&  0.87& 2.1\\
0.851& 0.17&  75&   510&  146&  1.00& 2.1\\
0.871& 0.15&  75&   599&  146&  1.17& 2.1\\
0.891& 0.13&  75&   726&  146&  1.42& 2.1\\
0.913& 0.11&  75&   920&  146&  1.80& 2.1\\
0.935& 0.08&  75&  1257&  146&  2.45& 2.1\\
0.958& 0.05&  75&  1981&  146&  3.87& 2.1\\
0.982& 0.02&  75&  4673&  146&  9.13& 2.1\\
0.988& 0.02&  75&  6875&  146& 13.43& 2.1\\
\hline
\end{tabular}
\end{center}
\footnotesize
NOTES.---%
As a function of the scale factor at the middle of each tile:
the weight, $t(1-t)$, at the tile midpoint, the box size used for that tile
and the size of the force mesh, the angular size of the box and mesh and
the particle mass.
These numbers are for the $512^3$ simulations.  For the $256^3$ simulations
the mesh size should be doubled and the mass per particle increased by $2^3$.
\end{table*}

\begin{figure}
\begin{center}
\leavevmode
\epsfxsize=3.5in \epsfbox{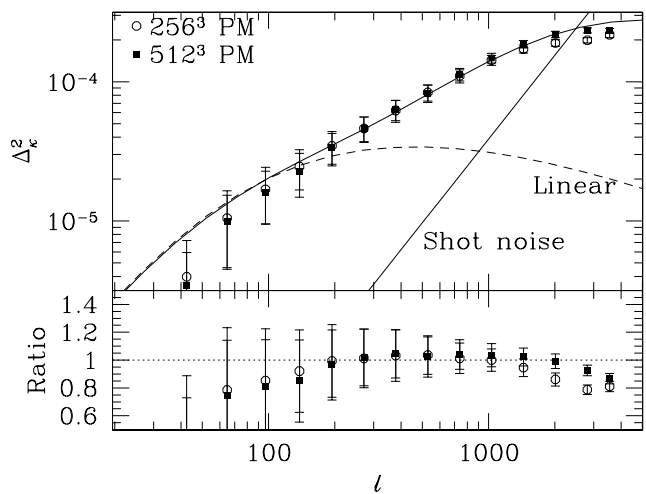}
\end{center}
\caption{\footnotesize%
(top) The angular power spectrum, $\ell^2\clk /(2\pi)$
or $\Delta_\kappa^2$, vs.~multipole number $\ell$, for the convergence
$\kappa$ from our tiling simulations.  We also show the semi-analytic
prediction from Eq.~(\protect\ref{eqn:semianalytic}) using both linear theory
and the non-linear power spectrum.  The shot-noise contribution assuming
$\bar n= 2\times 10^5$ galaxies per square degree each with an rms ellipticity
$\gamma_{\rm rms}=0.4$ is also shown.
(bottom) The ratio of the simulation results to the (non-linear) prediction of
Eq.~(\protect\ref{eqn:semianalytic}).}
\label{fig:pkap}
\end{figure}

\begin{figure}
\begin{center}
\leavevmode
\epsfxsize=3.5in \epsfbox{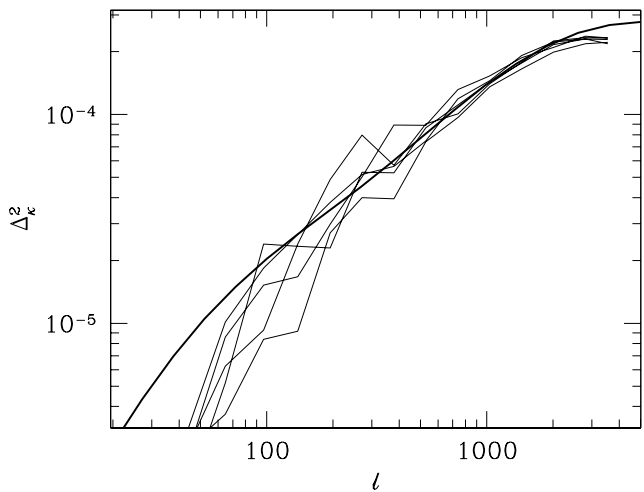}
\end{center}
\caption{\footnotesize%
As Fig.~\protect\ref{fig:pkap}, but with the power spectrum from
5 different realizations shown to emphasize the scatter from field-to-field.
The thick solid line is the prediction from
Eq.~(\protect\ref{eqn:semianalytic}).}
\label{fig:pkap_many}
\end{figure}

\section{Power Spectrum} \label{sec:power}

Fig.~\ref{fig:pkap} shows the angular power spectrum of $\kappa$, computed
{}from the tiling simulations, as compared to the semi-analytic result
(see Appendix),
\begin{eqnarray}
  \Delta_\kappa^2(\ell) &=& {9\pi\over 4\ell}
    \left[\Omega_m H_0^2 D_s^2\right]^2 \int {d D_{||}\over D_s}
    \ t^3 (1-t)^2 \nonumber\\
    && \quad\times
    \left[{\Delta_{\rm mass}^2(k=l/D_{||},a)\over a^2}\right] \,,
\label{eqn:semianalytic}
\end{eqnarray}
where $\Delta_{\rm mass}^2(k)=k^3P(k)/(2\pi^2)$
is the contribution to the mass variance per logarithmic interval 
physical wavenumber and analogously
$\Delta_\kappa^2(\ell)=\ell^2\clk/(2\pi)$ is the 
contribution to $\kappa_{\rm rms}^2$  per logarithmic interval
in angular wavenumber 
(or equivalently multipole) $\ell$.
We also show, in Fig.~\ref{fig:pkap_many}, the power spectrum from the first
5 realizations, to emphasize the scatter from field-to-field.

\begin{figure}
\begin{center}
\leavevmode
\epsfxsize=3.5in \epsfbox{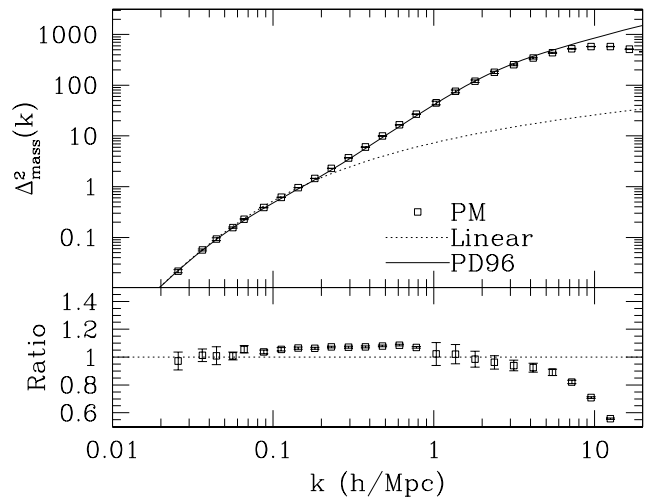}
\end{center}
\caption{\footnotesize%
(top) The 3D mass power spectrum from our ensemble of simulations as
compared to the fitting function of Peacock \& Dodds (\protect\cite{PeaDod96}).
Here {\it and only here\/} the error bars represents the error on the mean
computed from averaging over our many realizations of the model.
(bottom) The ratio of the N-body results to the fitting formula.}
\label{fig:pk}
\end{figure}

In evaluating Eq.~(\ref{eqn:semianalytic}), we use the method of
Peacock \& Dodds~(\cite{PeaDod96}) to compute the non-linear power spectrum
as a function of scale-factor.  Comparison with the average power spectrum
{}from our simulations (e.g.~Fig.~\ref{fig:pk} at $z=0$) shows agreement at
the 10\% level for the range of redshifts and scales resolved by
simulation ($k\simlt 5h$Mpc$^{-1}$).

The loss of power on large scales (small $\ell$) is a result of our FT based
analysis routines and the $6^\circ\times 6^\circ$ field of view.
To test this we generated gaussian fields with the angular power spectrum
of Eq.~(\ref{eqn:semianalytic}) and with much larger areas.
When analyzing $6^\circ\times 6^\circ$ subfields the same low-$\ell$
suppression as in Fig.~\ref{fig:pkap} was seen and comes from ``windowing''
the map by the field of view.

The roll-off at high-$\ell$ in Fig.~\ref{fig:pkap} is as expected from the
resolution of the N-body code.
The PM code resolves scales (in $k$) down to approximately $k_{\rm Nyquist}/3$
with a slight dependence on the spectral index of the model.
The smallest $k$ we can simulate is $2\pi/L_{\rm box}$ so in a $512^3$
simulation we would expect a dynamic range in $k$ of $256/3\simeq 90$.
The projection from physical scale to angular scale is not unique but rather
has a finite width ``kernel'' (see Eq.~\ref{eqn:semianalytic}).  In our case
the width is roughly a factor of 3 in scale.  So to fully resolve a given
$\ell$-mode, we need to resolve a factor of 3 higher in physical scale than
that mode projects at the mid-plane.  This further reduces our dynamic range
to a factor of 30.  Because we cannot arbitrarily reduce the box size without
the fundamental mode going non-linear by the present, our tiling is inefficient
at low-$z$ and our actual dynamic range is closer to a factor of 20--25, as
can be seen in Fig.~\ref{fig:pkap}.

\begin{figure}
\leavevmode
\begin{center}
\epsfxsize=3.5in \epsfbox{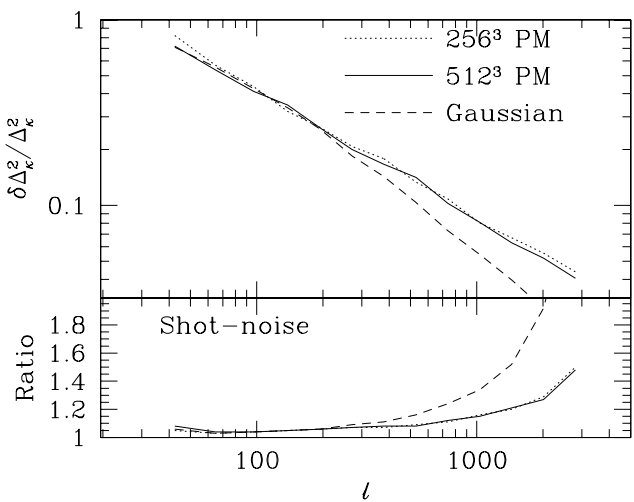}
\end{center}
\caption{\footnotesize%
(top) The standard deviation $\delta\Delta_\kappa^2$ in our binned
estimates of $\Delta^2_\kappa$ as a function of $\ell$.  The solid line
is from our $512^3$ simulations, dotted is our $256^3$ simulations and
the short-dashed line is from Gaussian fake skies with the same power spectrum.
The choice of binning is given in Table~\protect\ref{tab:cov}.
(bottom) The ratio of errors with shot-noise to pure sampling errors for the
three cases above.}
\label{fig:dpp}
\end{figure}

The error bars in Fig.~\ref{fig:pkap} are the sampling errors for an individual
$6^\circ\times 6^\circ$ field of view as estimated from the scatter of the full
suite of simulations.
{\it This should not be confused with the much smaller error on the mean power
spectrum of the suite.}
Sampling errors for different survey dimensions scale roughly as the ratio of
the dimensions and the variance as the ratio of the survey areas.

As demonstrated in Figs.~\ref{fig:pkap}, \ref{fig:dpp}, even though sampling
errors only fully converge to that of a gaussian random field with the same
power spectrum for $\ell\simlt 300$, the non-gaussian contribution to the
errors remains in the few tens of percent out until at least $\ell\simlt1000$
(in qualitative accord with analytic estimates, see
Scoccimarro, Zaldarriaga \& Hui~\cite{ScoZalHui}).
We have checked that the deviations from gaussianity are only weakly dependent
on the binning chosen for this range in $\ell$. 
This can also be seen be examining the covariance of the binned power spectrum
estimators shown in Table~\ref{tab:cov}. As with the variance, the covariance
deviates from the gaussian limit beginning at $\ell\sim 300$ and grows at
a moderate rate through $\ell\sim 1000$.   The bins are correlated even in 
the gaussian limit by the limited field of view: the fundamental mode implies
a spacing of $\Delta\ell=60$.  

\begin{figure}
\begin{center}
\leavevmode
\epsfxsize=3.5in \epsfbox{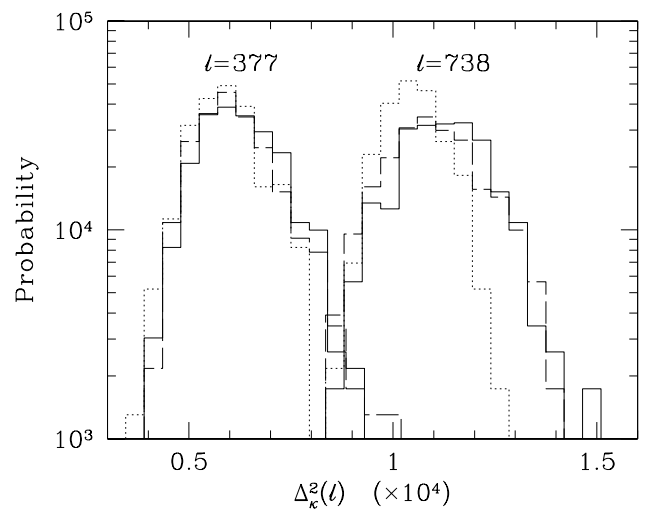}
\end{center}
\caption{\footnotesize%
The histogram of $\Delta^2_\kappa(\ell)$ for two different
bins from our higher (solid)
and lower (dashed) resolution simulations, and fake fields generated with
Gaussian statistics (dotted).}
\label{fig:pkap_hist}
\end{figure}

The full distribution of the power spectrum estimator also becomes moderately
less well characterized by its variance for $\ell \simgt 300$.
In Fig.~\ref{fig:pkap_hist}, we show the histogram of values from the
simulations.  The probability of outliers on the low side {\it decreases\/}
due to the skewness of the distribution whereas that on the high side
remains reasonably well characterized by the variance for $2$ and $3\sigma$
outliers.  These probabilities are with respect to a Gaussian sky artificially
set to the {\it same\/} variance for the power spectrum estimator.
This should not be confused with the expectations from the Gaussian prediction
for the variance: for the $\ell=738$ bin, a $>2\sigma$ deviation from the mean
power with respect to the Gaussian standard deviation occurs in one quarter
of our tiles.

Beyond $\ell\sim 1000$, the nongaussian contributions to the variance,
covariance, and tails of the distribution of the power spectrum estimators
becomes substantial.  
However, this is also the point at which shot noise from the intrinsic
ellipticity of the galaxies begins to exceed the sample variance. 
The shot noise power spectrum is (Kaiser \cite{Kai98})
\begin{equation}
  C_{\rm noise} = { \gamma_{\rm rms}^2 \over \bar n} \,,
\end{equation}
where $\bar n$ is the number density of the sources and $\gamma_{\rm rms}$ is
the rms intrinsic shear in each component.  
The shot noise spectrum for $\bar n = 2 \times 10^5 $deg$^{-2}$ and
$\gamma_{\rm rms}=0.4$ is shown in Fig.~\ref{fig:pkap}.
The noise bias in the measurements of the power spectrum can be subtracted off
at the expense of increasing the variance of the estimator for each $\ell$-mode
\begin{equation} 
\delta C_\ell^2 |_{\rm total}  = \delta C_\ell^2 |_{\kappa} 
	+ 4 C_\ell C_{\rm noise} + 2 C_{\rm noise}^2 \,.
\end{equation}	
For our binned estimators, the sample variance is reduced by $\sqrt{N}$
statistics so that the total fractional variance is
\begin{equation}
\left({\delta \clk \over \clk}\right)_{\rm total}^2 =
\left({\delta \clk \over \clk}\right)_{\rm sim}^2 +
        {1 \over N_\ell^2 \clk^2}  
	\sum_{\vec{\ell}} (4 \clk C_{\rm noise} + 2 C_{\rm noise}^2)
\label{eqn:powernoise}
\end{equation}
where the first term is the result from our simulations (without shot-noise)
and the sum in the second term is over the $N_\ell$ independent modes in the
bin.  The number of independent modes for a given $\ell$ is approximately
$(2\ell+1)f_{\rm sky}$, where $f_{\rm sky}$ is the fraction of sky covered
by the field of view ($f_{\rm sky}\sim 10^{-3}$ for our fields).
We show the effect of shot noise on the sample variance in Fig.~\ref{fig:dpp}.
We have tested these approximations against monte carlo realizations
of the shot noise and found good agreement. 

\begin{table*}
\begin{center}
\caption{\label{tab:cov}}
{\sc Power Spectrum Covariance\\}
\begin{tabular}{ccccccccccc}
\tablevspace{3pt}
\hline $\ell_{\rm bin}$   
       & 97      & 138     & 194     & 271     & 378     & 529     & 739   
  & 1031    & 1440   & 2012 \\
\hline
    97 & 1.00    & 0.26    & 0.12    & 0.10    & 0.02    & 0.10    & 0.12 
  & 0.15    & 0.18   & 0.19\\
   138 & (0.23)  & 1.00    & 0.31    & 0.21    & 0.09    & 0.14    & 0.16
  & 0.18    & 0.15   & 0.22\\
   194 & (0.04)  & (0.22)  & 1.00    & 0.26    & 0.24    & 0.28    & 0.17
  & 0.15    & 0.19   & 0.16\\
   271 & (-0.02) & (-0.03) & (0.17)  & 1.00    & 0.38    & 0.33    & 0.34
  & 0.27    & 0.19   & 0.32\\
   378 & (-0.01) & (0.02)  & (0.04)  & (0.11)  & 1.00    & 0.45    & 0.38
  & 0.33    & 0.32   & 0.27\\
   529 & (0.01)  & (-0.01) & (-0.07) & (-0.02) & (0.02)  & 1.00    & 0.50
  & 0.48    & 0.36   & 0.46\\
   739 & (0.04)  & (-0.03) & (-0.01) & (-0.02) & (-0.02) & (0.13)  & 1.00
  & 0.54    & 0.53   & 0.50\\
  1031 & (0.07)  & (0.01)  & (0.07)  & (-0.03) & (0.03)  & (0.08)  & (0.04)
  & 1.00    & 0.57   & 0.61\\
  1440 & (-0.03) & (0.02)  & (0.04)  & (-0.04) & (0.05)  & (-0.07) & (-0.04)
  & (-0.03) & 1.00   & 0.65\\
  2012 & (-0.02) & (-0.04) & (0.03)  & (0.03)  & (0.03)  & (-0.02) & (0.03)
  & (-0.07) & (0.02) & 1.00\\
\hline
\end{tabular}
\end{center}
\footnotesize
NOTES.---%
Covariance of the binned power spectrum estimators.   Upper triangle
displays the covariance found in the 512 tilings of the $512^3$ simulations.
Lower triangle (parenthetical numbers) displays the covariance found in an
equal number of gaussian realizations.  The finite $6^\circ \times 6^\circ$
field of view couples the power spectrum estimators over $\Delta\ell\sim 60$
in both cases, whereas non-linear dynamics couples the estimators in the
simulations at high $\ell$.
\end{table*}

The combination of these results imply that techniques based on gaussian 
assumptions for power spectrum estimation are  fair approximations
at least in the context of this $\Lambda$CDM model 
(e.g.~Kaiser \cite{Kai98}; Seljak \cite{Sel98}; Hu \& Tegmark \cite{HuTeg99}).

\section{Skewness and Kurtosis} \label{sec:skewness}

\begin{figure}
\begin{center}
\leavevmode
\epsfxsize=3.5in \epsfbox{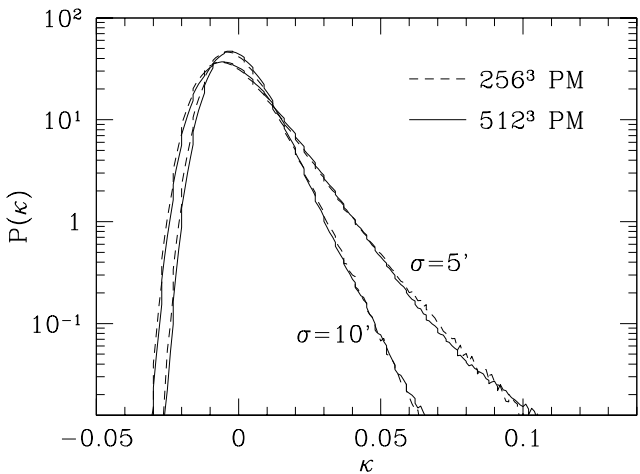}
\end{center}
\caption{\footnotesize%
The histogram of $\kappa$, smoothed with a top-hat filter of
radius $5'$ and $10'$.  The solid lines are from our $512^3$ simulations
while the dashed lines are from our $256^3$ simulations.}
\label{fig:histogram}
\end{figure}

\begin{figure}
\begin{center}
\leavevmode
\epsfxsize=3.5in \epsfbox{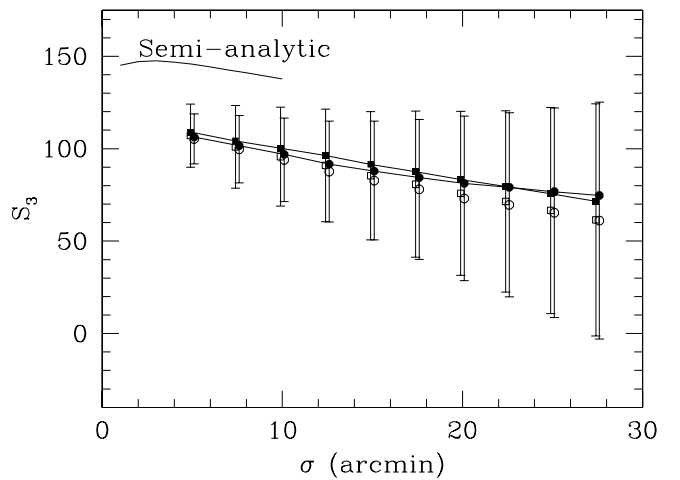}
\end{center}
\caption{\footnotesize%
The skewness, $S_3$, as a function of (top hat) smoothing scale.
The squares are results from the $512^3$ simulations while the circles are from
the $256^3$ simulations.  Filled symbols indicate the skewness computed from
the set of generated $\kappa$ planes while open symbols with error bars
indicate the mean and {\it variance\/} of $S_3$ for each plane.
The points are offset slightly for clarity.
The solid line is a semi-analytic estimate (L. Hui, private communication),
discussed in more detail in the text.}
\label{fig:s3}
\end{figure}

Figure~\ref{fig:histogram} shows the co-added histogram of $\kappa$, smoothed
on $5'$ and $10'$, from 512 tiling solutions.  The non-gaussian nature of the
distribution is apparent in this figure, as is the low-$\kappa$ cutoff
enforced by $\delta_{\rm mass}\ge -1$.
The large number of tiling solutions we have run allows us to probe the
distribution well into the tails.  Clearly the higher and lower resolution
simulations agree well on these scales.
Our ability to simulate many $\kappa$ planes allows us to study the statistics
of the moments of this distribution.  In this section, we examine the lowest
order moments beyond the 2-point function: the skewness and kurtosis.

\subsection{Simulation Results}

{}From the two dimensional angular grid of the convergence $\kappa$, we 
calculate the skewness and kurtosis on an angular scale $\smooth$.
We first smooth the grid with a pixelized tophat window $W_\smooth$ with
FT techniques
\begin{equation}
  \tilde\kappa_\smooth = \tilde \kappa \tilde W_\smooth
\label{eqn:2dsmth}
\end{equation}
and eliminate edge effects by zero padding the array and discarding
the data that is convolved with the zero padded region.  
We then calculate the skewness
\begin{equation}
S_3(\smooth) ={ \left< \kappa_\smooth^3 \right> \over 
		\left< \kappa_\smooth^2 \right>^2 }\,,
\label{eqn:s3}
\end{equation}
and the kurtosis
\begin{equation}
S_4(\smooth) =
   { \left\langle\kappa_\smooth^4\right\rangle -
    3\left\langle\kappa_\smooth^2\right\rangle^2
\over\left\langle\kappa_\smooth^2\right\rangle^3 }\,,
\label{eqn:s4}
\end{equation}
for two different averaging schemes:  averaging over pixels in a given 
$6^\circ \times 6^\circ$ field and averaging the pixels over all fields.  

\begin{figure}
\begin{center}
\leavevmode
\epsfxsize=3.5in \epsfbox{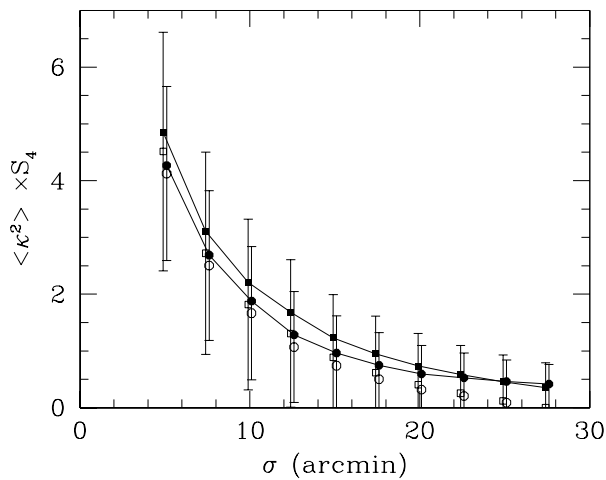}
\end{center}
\caption{\footnotesize%
As Fig.~\protect\ref{fig:s3} but for the kurtosis, $S_4$.
The kurtosis has been scaled by $\langle\kappa^2\rangle$ for display
purposes.}
\label{fig:s4}
\end{figure}

As can be seen in Figs.~\ref{fig:s3} and \ref{fig:s4}, even a
$6^\circ\times6^\circ$ field suffers from large sample variance on scales
of tens of arcminutes.
Like the power spectrum estimators, we expect the sample variance to scale
roughly with the survey area.
Through generating Gaussian fields with the same power spectrum, we find
that the sampling errors for $S_3$ and $S_4$ are a factor of 2 and 7 
larger than the Gaussian limit respectively for $\smooth\sim 10'$. 

The difference in the moments computed from averaging $S_N$ in each field
compared to computing $S_N$ using all of the field has been stressed by
Hui \& Gaztanaga~(\cite{HuiGaz}).    
The bias increases as the  field-to-field variance increases as can be
seen by comparing large and small smoothing scales in Figs.~\ref{fig:s3} 
and \ref{fig:s4}.
(In our simulations we found that the value of $S_N$ computed using the moments
of all the fields fluctuated more with increasing numbers of runs than the
mean of the $S_N$ computed from moments within each field.)
This large sampling errors should be borne in mind when employing $S_3$
measurements to distinguish between cosmological model.

Comparison of the $512^3$ simulations with the $256^3$ simulations indicates
that the N-body calculation has converged on a scale of $10'$, both in the
moments themselves and in the sampling errors.
The two sets of simulations begin to diverge in their fractional standard
deviation near $5'$, suggesting that the higher resolution simulations may
even be reliable down to $2.5'$.  Fig.~\ref{fig:s3_hist} shows the divergence
between the simulations in $S_3$ is in the high $S_3$ tail, which may
be due to resolution or may indicate too few higher resolution simulations
have been run.
We have also checked that the $75 h^{-1}$Mpc are large enough to provide an
adequate sample of the non-linear scale for these purposes.
Omiting these simulations and completing the tiling with $95 h^{-1}$Mpc
simulations produces a negligible change in $S_3$ at $10'$.

As Fig.~\ref{fig:s4} shows, the kurtosis increases above 
the 
$\langle\kappa^2\rangle S_4=3$ below $10'$. 
As this is the number expected for $\left< \kappa_\sigma^4 \right>/
\left< \kappa_\sigma^2 \right>^3$ for a gaussian field,
it marks the regime where the distribution becomes significantly
non-gaussian in the 4th moments.  
However we
detect no similar dramatic 
rise in the power spectrum errors at $\ell \sim 1000$
(\S\ref{sec:power}).

Finally we have simulated the effect of shot-noise on the variance of $S_3$
and $S_4$.  In the presence of shot-noise we define estimators of $S_N$ in
analogy with Eqs.~(\ref{eqn:s3}, \ref{eqn:s4}) but which subtract the
contribution of the shot-noise to $\langle\kappa_\smooth^n\rangle$.
For example if $\kappa_\smooth'$ is the {\it measured\/} value of
$\kappa_\smooth$ including shot-noise with variance
$\langle\epsilon_\smooth^2\rangle$, we define
\begin{equation}
  S_3 = {\left\langle{\kappa_\smooth'}^3\right\rangle\over
         \left\langle{\kappa_\smooth'}^2\right\rangle-
         \left\langle\epsilon_\smooth^2\right\rangle}
\end{equation}
Using these estimators and adding simulated shot-noise to our planes we find
that the estimators are unbiased and their standard deviations are only
slightly increased ($\simlt 16\%$ for $S_3$ and $\simlt 6\%$ for $S_4$) even
on scales as small as $2.5'$.
This is not too surprising since with $2\times 10^5$ galaxies per square degree
the shot-noise power only surpasses the signal power in our model on scales
smaller than $1.3'$, and we have shown that the sample variance on $S_3$ and
$S_4$ is significantly enhanced by the non-Gaussianity of the distribution.
Artificially increasing the noise by a factor of 4 does lead to an increase
in the variance of $S_3$ and $S_4$, but the estimators remain unbiased.

\subsection{Comparison to Previous Results}

\begin{figure}
\begin{center}
\leavevmode
\epsfxsize=3.5in \epsfbox{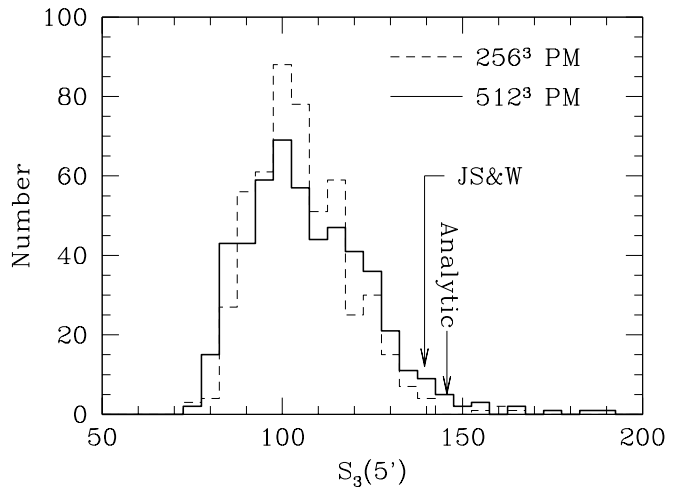}
\end{center}
\caption{\footnotesize%
The histogram of $S_3(5')$ values from our higher (solid) and
lower (dashed) resolution simulations.  Also shown are the predictions
{}from HEPT (see text) and one plane of the simulations of
Jain et al.~(\protect\cite{JaiSelWhi98}).}
\label{fig:s3_hist}
\end{figure}

These results make sense physically, but it would be useful to compare with
previous work.
On the scales we are working perturbation theory is not adequate, so the best
comparison is with other simulations, the closest being those of
Jain, et al.~(\cite{JaiSelWhi98}).
Unfortunately a direct comparison with their work is difficult.
While our model differs slightly from theirs, we have run a smaller set of
simulations of their exact model and find that $S_3$ are not strongly affected
by the slight changes.
However we do not have the dynamic range to reliably estimate the skewness
on $1'$ scales, and their $3.5^\circ\times3.5^\circ$ field is sufficiently
small that they have large sample variance on $10'$ scales.
Using our analysis software on one field from their simulations
(B. Jain, private communication), our skewness is approximately 20\% lower
at $5'$ than theirs.
We compare at $5'$, which is the edge of our reliable range, because the sample
variance from their small fields makes comparison difficult above this scale.
Indeed, in the plane we have analyzed their skewness peaks at $10'$ before
dropping precipitously.
In the distribution of $S_3$ in our higher resolution ($512^3$ mesh)
simulations (see Fig.~\ref{fig:s3_hist}) only 5\% of our planes have $S_3$
as high or higher than the plane from Jain et al.~(\cite{JaiSelWhi98}).
Crudely accounting for the increased sample variance due to their smaller
field by scaling the distribution by $6^\circ/3.5^\circ$, raises this number
to $10-15\%$.
While this is not highly improbable, the difference may still be due to
systematic differences in the codes.  Jain, et al.~(\cite{JaiSelWhi98}) also
performed some PM runs in a $64h^{-1}$Mpc box with a $256^3$ force mesh and
found results $\sim 20\%$ lower than their P${}^3$M results at $5'$
(Seljak, private communication).  Whether this discrepancy is due to the small
box size they used, systematic difference between PM and P${}^3$M
(e.g.~Splinter et al.~\cite{SplMelShaSut};
Jain \& Bertschinger\footnote{The discrepancy noted by these authors for the
$n=-2$ spectrum is not directly relevant here, since $n$ is less negative on
the scales of interest.  We have shown this specifically in
Fig.~\protect\ref{fig:pk}.  The general point remains valid
however.}~\cite{JaiBer}) or sample variance is not clear.

\begin{figure}
\begin{center}
\leavevmode
\epsfxsize=3.5in \epsfbox{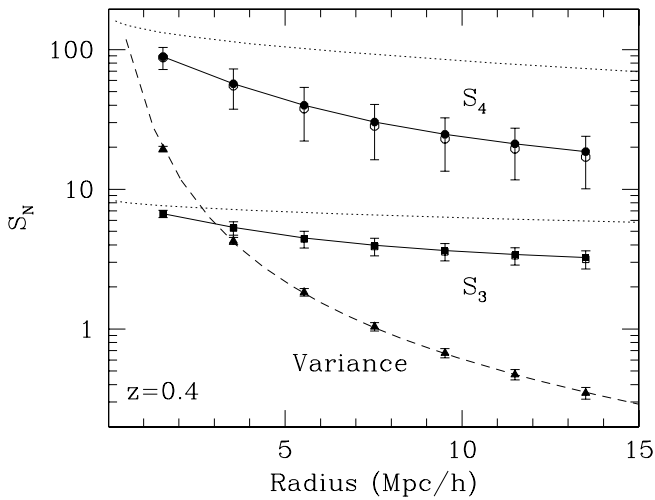}
\end{center}
\caption{\footnotesize%
The variance, skewness, $S_3$, and kurtosis, $S_4$, of the density
field at $z=0.4$ as a function of (top hat) smoothing scale.
These results are from our $512^3$ simulations, as described in the text.
Filled symbols indicate the $S_N$ computed from 10 boxes, while open symbols
with error bars indicate the mean and {\it variance\/} of $S_N$ for each box.
The dotted lines are the predictions based on HEPT, used to generate the
curve in Fig.~\protect\ref{fig:s3}.  The dashed line is the prediction of
Peacock \& Dodds~(\protect\cite{PeaDod96}).}
\label{fig:den_mom}
\end{figure}

We show in Figs.~\ref{fig:s3}, \ref{fig:s3_hist} the prediction of a
semi-analytic calculation (L. Hui, private communication) based on
hyper-extended perturbation theory
(HEPT; Scoccimarro \& Frieman~\cite{ScoFri}).
The agreement is at the level one would expect from the approximation used.
To check this we have calculated the skewness and kurtosis of the density
field (at $z=0.4$, the peak of the curves in Fig.~\ref{fig:dclda}) in
our $155h^{-1}$Mpc boxes with $256^3$ particles and $512^3$ force mesh.
For each of the 10 simulations, we binned the particles onto a $512^3$ grid
using NGP assignment (the results do not depend on this choice), then smoothed
this grid using a 3D analogue of Eq.~(\ref{eqn:2dsmth}) with a top hat 
radius $R$.
The moments were computed by averaging powers $\delta$ over the $512^3$ grid
sites.  Again we computed the average $S_N$ over the simulations and the
``global'' $S_N$ from combining the moments from all the simulations:
\begin{eqnarray}
S_3(R) & \equiv & {\left\langle\delta_R^3\right\rangle\over
                \left\langle\delta_R^2\right\rangle^2}\,, \\
S_4(R) & \equiv & {\left\langle\delta_R^4\right\rangle-
               3\left\langle\delta_R^2\right\rangle^2\over
                \left\langle\delta_R^2\right\rangle^3}\,.
\end{eqnarray}
Our results are shown in Fig.~\ref{fig:den_mom} as a function of radius,
along with the variance in the density.  Also plotted (dotted) are the
predictions of HEPT as used in the semi-analytic lensing calculation
(Hui~\cite{Hui}) and the variance (dashed) predicted by
Peacock \& Dodds~(\cite{PeaDod96}).
The non-gaussianity in the lensing signal may be generated at lower $z$ than
the peak in Fig.~\ref{fig:dclda}, so we have also calculated $S_3$ and $S_4$
{}from our $z=0$ data.  The results are consistent with little or no evolution
in $S_3$ and $S_4$ since $z=0.4$, though the variance grows as predicted by
Peacock \& Dodds~(\cite{PeaDod96}).

The level of disagreement is sufficient to explain the discrepancy in
Fig.~\ref{fig:s3}.  The stated realm of validity for the HEPT
result is for variances $\simgt 100$ and indeed our results
for $S_3$ and $S_4$ of the density
in this regime agree better with the prediction 
(see Fig.~\ref{fig:den_mom}).  For lower variances, one expects
both moments to be smaller and the power law approximation to
the mass power spectrum inherent in HEPT to break
down.  Preliminary results from a treatment that includes
these two effects (R. Scoccimarro, private communication) indeed 
agree better with our lensing results: $S_3 \approx 115$ at $5'$, compared
with our $111$, with a gradual decline to $S_3 \sim 80$ at $70'-80'$
before an increase back to the perturbation theory results
of Bernardeau et al.\ (\cite{BerWaeMel97}).
In any case, the difference shown in Fig.~\ref{fig:den_mom} is 
very likely the cause of the discrepancy with the 
semi-analytic calculation.


\section{Tiling Tests} \label{sec:singlebox}

Due to our ``tiling'' method, and the large number of simulations that we
have run (Table~\ref{tab:numsim}), we are able to systematically examine the
dependence of our various results on the volume of space sampled by the
simulations.  Of particular interest is the following question: how is
the sample variance associated with each field affected by tracing repeatedly
through a single simulation?
We have attempted to answer that question for various statistics with our
large ensemble of simulations.

We first looked at the statistics of the power spectrum.
Using our 76 large boxes ($245h^{-1}$Mpc at $256^3$ force resolution) we
checked that the mean, variance, and pdf of the power spectrum (for 4 different
binnings) was the same whether we shuffled the tiles between boxes or used
each box in isolation.  These large and lower-resolution boxes are not fully
resolving the structure at late times, but this is not of great concern as the
small scale structure that is missing is unlikely to be correlated over
large scales as required to cause an effect in this test.

For the power spectrum, repeatedly tracing through a $245h^{-1}$Mpc box
provides the same distribution as our tiling method (though multiple boxes
are still needed to assess sample variance).
The same test for our smaller $75h^{-1}$Mpc also shows no statistically
significant effect. This is significant since as the box shrinks the 
volume of space sampled is reduced and sampling
becomes a larger issue.  
Likewise for $S_3$ and $S_4$ no significant difference between repeated
tracing and tiling was found.

These results shed light on another possible concern: that the rays in these
simulations trace through boxes which are joined ``discontinuously'' at their
edges.  Fig.~\ref{fig:pkap} shows that this does not affect the mean value of
the 2-point function reproduced by the code.
Our multiple simulations allow us to go further however.
A comparison of the tiling simulations with the repeated tracings of the
$245h^{-1}$Mpc box allows us to test the effect of a different number of
``matchings'' along the line-of-sight.
On the scales where both resolve the structure, we find convergence in the
mean value of $S_3$ (the values of $S_4$ are too noisy to allow a strong
statement) and the pdf of $\Delta_\kappa^2$ with 4 different binnings.
This is suggestive that box matching is not a major source of error, though
we cannot test that this is true on very small scales due to lack of
resolution in our simulations.

\begin{figure}
\begin{center}
\leavevmode
\epsfxsize=3.5in \epsfbox{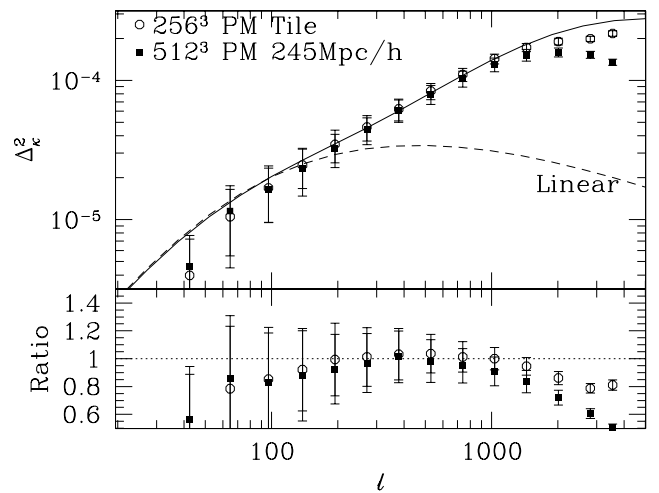}
\end{center}
\caption{\footnotesize%
Comparison of resolution from tiling compared with single size boxes.
The open circles represent the power spectrum of fields produced by ``tiling''
with the $256^3$ PM simulations.
The filled squares are the power spectrum deduced from our higher resolution,
$512^3$ simulation, but using only the largest box: $245h^{-1}$Mpc.
Notice that tiling wins back the extra factor of 2 in resolution.}
\label{fig:snglcmp}
\end{figure}

Tiling does increase the angular resolution of our simulations.
In Fig.~\ref{fig:snglcmp} we compare the angular power spectrum from our
higher resolution ($512^3$ mesh) PM runs in boxes of size $245h^{-1}$Mpc
with our tiling solution at lower resolution ($256^3$ mesh), using the
shrinking box.
One can see that allowing the box to shrink along the line-of-sight
produces considerable gains in angular resolution.  Unfortunately the need
to keep the fundamental mode of the box linear at all times restricts the
size of the low-$z$ boxes and limits the gain in angular resolution which
can be achieved by this method (to a factor between 2 and 3).
While larger fields of view are easily simulated, the minimum size of the
low-$z$ boxes restricts the smallest angular scale that can be probed.
This gain is enough, however, to make PM codes viable for a rapid exploration
of parameter space on workstation class machines
(cf.~Jain, et al.~\cite{JaiSelWhi98}).

\section{Discussion} \label{sec:discussion}

We have described an efficient algorithm for calculating the statistics of
weak lensing by large-scale structure in N-body simulations.  By working
with the unperturbed paths, our method is extremely simple to implement
and can be done at the same time as the N-body run(s).
This gives one the ability to simulate a large volume and sample the
line-of-sight integration densely in both space and time.
Contrast this with more traditional ray tracing techniques which use
only tens of lens planes and project the entire density distribution in the
box onto a single lens plane for each time step.
Neglecting the deflections is certainly self-consistent within the weak lensing
approximation.
Analytic arguments (Bernardeau et al.~\cite{BerWaeMel97}) and explicit ray-tracing
simulations (Jain, et al.\ \cite{JaiSelWhi98}) furthermore imply that
corrections due to deflections are small for our purposes.

As with other simulation based results, numerical resolution and dynamic
range are a serious issue.  In particular the effects of finite force
resolution can be seen in our results below $2.5'$.
For weak lensing there are also problems introduced by the periodicity of
the simulation box, which limits the size of the field of view that one can
probe in a given simulation, in our case to $6^\circ\times6^\circ$.

We have described a technique, which we call ``tiling'', which allows us to
use results from multiple realizations of a given model, and to match the size
of the simulation box to the converging ray bundle to increase angular
resolution for a fixed physical resolution.  By varying the tiling
scheme, we also tested the effects of discontinuities from joining
the boxes and repeatingly tracing through the same simulation.  
We found no significant effect from either. 

With our suite of simulations, we are able to predict not just the mean
properties of the models, but also their sampling errors.  
This is extremely important in assessing the statistical significance of
future measurements.
We have shown that the non-gaussian contribution to the errors on the power
spectrum remains small out to $\ell\sim 10^3$, even though the distribution
of convergence, $\kappa$, is clearly non-gaussian at $10'$.
We have quantified the (large) sample variance in estimates of higher moments
of the $\kappa$ distribution which have been suggested as tests of the energy
contents of the universe.
Even with $6^\circ\times6^\circ$ fields the errors on the moments are totally
dominated by sample variance on scales above a few arcminutes with galaxy
densities achievable in current observations.

Since sampling errors scale inversely with the dimensions of the survey,
a field of view in the tens to hundreds of square degrees will be crucial
for extracting cosmological information on large scale structure from weak
lensing surveys, especially for the non-gaussian signatures of models.
Nevertheless, due to the growing number of instruments with wide fields of
view, for example MEGACAM at CFHT (Boulade et al. \cite{Bou98}) and the VLT
Survey Telescope at the European Southern Observatory
(Arnaboldi et al. \cite{Arn98}), the prospects for weak lensing in the era
of precision cosmology remain bright.

\bigskip
\acknowledgments  
We would like to acknowledge useful conversations with
R. Barkana, J. Cohn, R. Croft, L. Hui, B. Jain, 
R. Scoccimarro, U. Seljak \& M. Zaldarriaga.
M.W. thanks J. Bagla for numerous helpful conversations on N-body codes.
W.H. was supported by NSF-9513835, the Keck Foundation, and a Sloan Fellowship.
M.W. was supported by NSF-9802362.
Parts of this work were done on the Origin2000 system at the National
Center for Supercomputing Applications, University of Illinois,
Urbana-Champaign.

\appendix

\section{Limber's equation}

In this appendix we provide a simple derivation of the expression in the
main text for the 2-point function of the convergence, $\kappa$.
We start by assuming that the lensing occurs at late enough times
that the anisotropic stress of the radiation 
can be neglected, so that in Newtonian gauge
we can write the metric
\begin{equation}
ds^2 = a^2(\eta)\left[ -(1+2\Phi)d\eta^2 + (1-2\Phi) d\vec{x}^2 \right]
\end{equation}
to first order in the gravitational potential $\Phi\sim 10^{-5}$.
Here $d\eta=a(t)dt$ is the conformal time and we have written the 3-metric
schematically as $d\vec{x}^2=d\chi^2+r^2(\chi) d\Omega$.
For scales smaller than the curvature scale we can approximate this as flat,
$r(\chi)=\chi$, however on cosmological scales we need to use
$r=|K|^{-1/2}\sinh|K|^{1/2}\chi$ for an open universe and $r=|K|^{-1/2}\sin|K|^{1/2}
\chi$ for a closed universe.
The conformal factor, $a(\eta)$, accounts for the cosmological redshift of
photon energy.  When following null geodesics we may scale it out,
i.e.~set $a=1$.  The Lagrangian describing geodesic motion is
$L={1\over 2}g_{\mu\nu}\dot{x}^\mu\dot{x}^\nu$ where overdot represents
differentiation with respect to an affine parameter $\lambda$ along the path.
Recalling that the momentum $p_\perp\equiv dL/d\dot{x}_\perp$ the
Euler-Lagrange equations become
\begin{equation}
  {dp_\perp\over d\lambda} = {\partial L\over\partial x_\perp}
  = -2 {\partial\Phi\over\partial x_\perp} p_{||} {dx_{||}\over d\lambda}
  + {\cal O}(\Phi^2)\,.
\end{equation}
Thus the deflection angle, $\alpha\equiv\Delta p_\perp/p_{||}$, receives
differential contribution $d\alpha=-2\nabla_\perp\Phi dx_{||}$.
Simple geometry dictates that such a deflection at a ``lens'' position
results in a change of angle at the observer of
$\delta\theta=(r_{LS}/r_S)\delta\alpha$ where $r_{LS}=r(\chi_S-\chi_L)$ is
the (radial) distance from the lens to the source and $r_S=r(\chi_S)$ is
the (radial) distance from the observer to the source.
Translating this into a change in position at $\chi_L$ of
$\delta x_\perp=D_L\delta\theta$ we see that two rays initially seperated
by $D_S\Delta\theta$ have a final seperation
\begin{equation}
  \Delta x_i = \left( \delta_{ij} - \psi_{ij} \right) D_S\Delta\theta_j\,,
\end{equation}
where
\begin{equation}
  \psi_{ij} = 2\int_0^{\chi_S} d\chi\ {r_L r_{LS}\over r_S}
              \ \partial_i\partial_j\Phi\,.
\label{eqn:psidef}
\end{equation}
The $2\times2$ matrix $\left( \delta_{ij} - \psi_{ij} \right)$
can be expanded in Pauli matrices with coefficients
\begin{equation}
  \left( \delta_{ij} - \psi_{ij} \right) =
    (1-\kappa) I - \gamma_1 \sigma_3 - \gamma_2 
\sigma_1 \,,
\end{equation}
so e.g.~$\kappa={1\over 2}{\rm Tr}( I
\psi)={1\over 2}\psi_{jj}$ which leads
to Eq.~(\ref{eqn:losint}).
Since all of the coefficients are derived from one function, specifying any
one of them is sufficient.  We shall focus here on the convergence, $\kappa$.
Replacing $\nabla_\perp^2$ with $\nabla^2$ in the integral results in errors
of ${\cal O}(\Phi)\sim 10^{-5}$, so we can use
$\nabla^2\Phi=4\pi G\rho a^2\delta$ to obtain Eq.~(\ref{eqn:losintdelt}).

To calculate the 2-point function of $\kappa$ we expand $\delta(x)$ in
Fourier modes and use the Rayleigh expansion of a plane wave to find
\begin{eqnarray}
  \clk &=& 4\pi \left[ {3\over 2}\Omega_m H_0^2 \right]^2
          \int {dk\over k} \Delta_{\rm mass}^2(k) 
          \int d\chi_1 
          \int d\chi_2 
	\nonumber\\
         &&\quad  \times
		\left[ 
		{g(\chi_1)\over a_1} 
		{g(\chi_2)\over a_2}\right] 
		j_\ell(k\chi_1) j_\ell(k\chi_2)
\end{eqnarray}
where $g(\chi)$ is the distance kernel in Eq.~(\ref{eqn:psidef}).
For open universes replace $j_\ell$ with the hyperspherical bessel
function.
On small scales we may use the resolution of the identity
\begin{equation}
  \int k^2dk\ j_\ell(k\chi_1)j_\ell(k\chi_2)
  = {\pi\over 2} [r(\chi)]^{-2} \delta(\chi_1-\chi_2)\,,
\end{equation}
to obtain the power per logarithmic interval in $\ell$,
\begin{equation}
  {\ell(2\ell+1)\clk\over 4\pi} \simeq
  {9\pi\over 4\ell} \left[ \Omega_m H_0^2 \right]^2
  \int d\chi\ r  \left[ {g(\chi)\over a} \right]^2
  \Delta_{\rm mass}^2(\ell/r)\,.
\end{equation}
In a spatially flat universe ($r=\chi$), this reduces to
Eq.~(\ref{eqn:semianalytic}).


\begin{thebibliography}{99}
\bibitem[1998]{Arn98}
Arnaboldi M. et al. 1998, in Wide Field Surveys in Cosmology,
        ed. S. Colombi, Y. Mellier, \& B. Raban, (Paris: 
	Editions Frontieres), 343
\bibitem[1986]{BBKS}
Bardeen J.M., Bond J.R., Kaiser N., Szalay A. S., 1986, \apj, 304, 15
\bibitem[2000]{BarSch}
Bartelmann M., Schneider P., Phys. Rep., in press [astro-ph/9912508]
\bibitem[1997]{BerWaeMel97}
Bernardeau F., van Waerbeke L., Mellier Y. 1997,
	A\&A 322, 1
\bibitem[1991]{Blaetal91}
Blandford R.D., Saust A.B., Brainerd T.G., Villumsen J.V.,
  1991, MNRAS, 251, 600
\bibitem[1998]{Bou98}
Boulade O. et al.\ 1998, Proc. SPIE, 3355, 614
\bibitem[1997]{BunWhi}
Bunn E., White M., 1997, \apj, 480, 6
\bibitem[1998]{CouBarTho98}
Couchman H.P.M., Barber A.J, Thomas P.A., 1999, MNRAS, 310, 453 [astro-ph/9810063]
\bibitem[1998]{EkeColFreHen}
Eke V., Cole, S., Frenk, C.S., Patrick, H.J. 1998, \mnras, 298, 1145
\bibitem[1998]{FluWebMor98}
Fluke C.J., Webster R.L., Mortlock D.J., 1999, MNRAS, 306, 567
[astro-ph/9812300] 
\bibitem[1999]{HamMarFut99}
Hamana T., Martel H., Futamase T., preprint [astro-ph/9903002]
\bibitem[1981]{HocEas}
Hockney R.W., Eastwood J.W., 1981, Computer Simulation using Particles
(New York; McGraw-Hill)
\bibitem[1999]{HuTeg99}
Hu W., Tegmark M., 1999, ApJ,  514, 65 [astro-ph/9811168] 
\bibitem[1999]{Hui}
Hui L., 1999, ApJ, 519, 9 [astro-ph/9902275]
\bibitem[1999]{HuiGaz}
Hui L., Gaztanaga E., 1999, ApJ, 519, 622 [astro-ph/9810194]
\bibitem[1998]{JaiBer}
Jain B., Bertschinger E., 1998, \apj, 509, 517
\bibitem[1997]{JaiSel97}
Jain B., Seljak U., 1997, ApJ, 484, 560 [astro-ph/9611077]
\bibitem[1999]{JaiSelWhi98}
Jain B., Seljak U., White S.D.M., 1998, preprint [astro-ph/9804238]
\bibitem[1992]{Kai92}
Kaiser N., 1992, \apj, 388, 272
\bibitem[1998]{Kai98}
Kaiser N., 1998, \apj, 498, 26 [astro-ph/9610120]
\bibitem[1999]{MWP}
Meiksin A., White M., Peacock J. 1999, MNRAS 304, 851 [astro-ph/9812214]
\bibitem[1998]{Mellier}
Mellier Y., 1998, Ann. Rev. Astron. Astrophys., 37, 127 [astro-ph/9812172]
\bibitem[1991]{Mir91}
Miralda-Escude J., 1991, ApJ, 380, 1
\bibitem[1996]{PeaDod96}
Peacock J. A., Dodds S. J., 1996, MNRAS, 280, 19
\bibitem[1999]{ScoFri}
Scoccimarro R., Frieman J.A., 1999, ApJ, 520, 35 
\bibitem[1999]{ScoZalHui}
Scoccimarro R., Zaldarriaga M., Hui L., 1999, ApJ, 527, 1 [astro-ph/9901099]
\bibitem[1998]{Sel98}
Seljak U., 1998, ApJ, 506, 64
\bibitem[1998]{SplMelShaSut}
Splinter R., et al., 1998, \apj, 497, 38
\bibitem[1999]{Ste}
Stebbins A., [astro-ph/9609149]
\bibitem[1999]{ViaLid}
Viana P.T.P., Liddle A.R., 1999, MNRAS, 303, 535 [astro-ph/9803224]
\bibitem[1998]{WamCenOst98}
Wambsganss J., Cen R., Ostriker J.P., 1998, \apj, 494, 29
\bibitem[1999]{W99}
White M., 1999, MNRAS, 310, 51
[astro-ph/9811227].
\bibitem[1998]{WaeBerMel98}
Waerbeke L.V., Bernardeau F., Mellier Y., 1999, A\&A, 342, 15  [astro-ph/9807007] 
\end{thebibliography}
\end{document}